\documentclass[runningheads]{llncs}
\usepackage{graphicx}
\usepackage{xcolor}
\usepackage{todonotes}
\usepackage{amsmath}
\usepackage{float}
\usepackage{amssymb}
\usepackage{array}
\usepackage{footmisc}
\usepackage{printlen}
\usepackage{cite}
\usepackage{xcolor}
\usepackage{ulem}
\usepackage{caption}
\usepackage{subcaption}
\usepackage{comment}
\usepackage{hyperref}

\begin{document}
\title{Unsupervised Diffeomorphic Surface Registration and Non-Linear Modelling}
%
%
\author{Balder Croquet\inst{1,2} \and 
Daan Christiaens\inst{1,2} \and 
Seth M. Weinberg\inst{4} \and 
Michael Bronstein\inst{5,6,7} \and 
Dirk Vandermeulen\inst{1,2} \and 
Peter Claes\inst{1,2,3}} 
%
\authorrunning{B. Croquet et al.}
%
\institute{
Medical Imaging Research Center, UZ Leuven, Leuven, Belgium\\ \email{balder.croquet@kuleuven.be}\and
Department of Electrical Engineering, ESAT/PSI, KU Leuven, Leuven, Belgium\and
Department of Human Genetics, KU Leuven, Leuven, Belgium\and
Center for Craniofacial and Dental Genetics, Department of Oral and Craniofacial Sciences, University of Pittsburgh, Pittsburgh, USA\and
Department of Computing, Imperial College London, UK\and
IDSIA, USI Lugano, Switzerland\and
Twitter, UK
}

\maketitle              
\begin{abstract}
Registration is an essential tool in image analysis. Deep learning based alternatives have recently become popular, achieving competitive performance at a faster speed. However, many contemporary techniques are limited to volumetric representations, despite increased popularity of 3D surface and shape data in medical image analysis. We propose a one-step registration model for 3D surfaces that internalises a lower dimensional probabilistic deformation model (PDM) using conditional variational autoencoders (CVAE). The deformations are constrained to be diffeomorphic using an exponentiation layer. The one-step registration model is benchmarked against iterative techniques, trading in a slightly lower performance in terms of shape fit for a higher compactness. We experiment with two distance metrics, Chamfer distance (CD) and Sinkhorn divergence (SD), as specific distance functions for surface data in real-world registration scenarios. The internalised deformation model is benchmarked against linear principal component analysis (PCA) achieving competitive results and improved generalisability from lower dimensions. 

\keywords{Diffeomorphic registration \and Geometric deep learning \and Deformation Modelling}
\end{abstract}
\section{Introduction}
3D surface scans are an increasingly popular low-cost radiation-free alternative or addition to traditional medical imaging modalities. In this work we focus on 3D surface scans of faces, which are for instance used in the clinic for surgical planning and auditing of craniofacial reconstructive surgery \cite{matthews2020pitfalls}, or in medical research to describe and understand the effects of teratogens on facial development \cite{muggli2017association}. Traditionally, analyses were executed using sparse landmarks, more recent studies make use of non-rigid registration to transfer thousands of corresponding landmarks from a template to a target \cite{matthews2020pitfalls}. MeshMonk \cite{white2019meshmonk} is an open-source toolbox that was developed and validated for phenotyping of faces. It consists of a scaled rigid alignment step, based on rigid Iterative closest point (ICP), followed by an iterative non-rigid registration step that makes use of a Visco elastic model. Its major limitation is that it is iterative, with a complexity that scales with the number of vertices. Furthermore, MeshMonk does not provide any theoretical guarantees on the deformation. Contrarily, diffeomorphic non-rigid registration, enforces the deformation to be smooth, differentiable, invertible and topology preserving \cite{ashburner2007fast}, thus, providing extra robustness in registration. Furthermore, the metric induced by the diffeomorphism can be used as an alternative over point-to-point distances. State of the art iterative techniques include DARTEL \cite{ashburner2007fast}, Diffeomorphic Demons \cite{vercauteren2009diffeomorphic} and LDDMM \cite{beg2005computing, brunn2021fast} on volumetric images, with LDDMM also being generalised to surfaces \cite{vaillant2005surface}. Deformetrica \cite{bone2018deformetrica} is an open-source implementation of a specific instance of LDDMM that makes use of control points, it is applicable to many representations. The main limitation of Deformetrica is that it is iterative and thus still not usable in real-time.

Recently, the success of deep learning (DL) in image tasks have sparked interest in DL based volume registration. Supervised approaches such as \cite{yang2016fast}, making use of generated ground truth data can result in bias, leading to the introduction of the first unsupervised approaches \cite{dalca2018unsupervised, krebs2018unsupervised}. For surface data such as point clouds or meshes, the DL literature is far less expanded. Rigid registration of point clouds has been demonstrated in robotics \cite{zhang2020deep}. Le Clerc and Sun \cite{le2020memory} proposed a convolutional neural network for a vertex classification based mesh registration. Fu et al. \cite{fu2020deformable} proposed a deep neural network that performs volumetric point-cloud registration of segmented prostates in a multi-modal setting using bio-mechanical constraints. Bahri et al. \cite{bahri2020shape} proposed an encoder-decoder architecture capable of non-rigid registration of faces by translation of latent geometric information. FlowNet3D \cite{liu2019flownet3d}, predicts scene flow directly from unstructured point clouds, it is trained with synthetic ground truth data. There are also a few diffeomorphic surface registration models. ResNet-LDDMM \cite{amor2021resnet} uses a deep residual neural network based on LDDMM to register 3D shapes. Dalca et al. \cite{dalca2019unsupervised}, introduced a probabilistic model for diffeomorphic registration of volumes that has the option to incorporate a surface deformation model. However, to the best of our knowledge, a standalone deep learning based diffeomorphic model in conjunction with a lower-dimensional probabilistic deformation model has not yet been explored for surfaces.

In this work we introduce a geometric conditional variational autoencoder (CVAE) network \cite{kingma2014semi} based on \cite{krebs2018unsupervised, dalca2019unsupervised}. The network is able to perform a one-step 3D surface registration by deforming the ambient space while jointly learning a lower dimensional probabilistic deformation model (PDM). Because we deform the ambient space instead of the individual vertices, the model is independent of the resolution and representation of the input. Futhermore, when trained, it could be used for real-time patient assessment. The registration component is benchmarked against an iterative registration technique in a real-world scenario. The PDM serves as a learned active shape model, imposing a probabilistic prior on the deformations. The PDM is benchmarked against a traditional deformation model on a dataset that is in correspondence. The source code is publicly available.\footnote[1]{\url{https://gitlab.kuleuven.be/u0132345/deepdiffeomorphicfaceregistration}}


\section{Methods}
\subsection{Registration Model} \label{subsec:reg_model}
Assuming that the moving shape $M$ and fixed shape $F$ are embedded in $\Omega = \mathbb{R}^d$, shape registration involves finding a deformation function $\varphi$ that can be directly applied such that the two shapes maximally overlap $F\approx \varphi \circ M$. We choose to work with a mapping $\varphi: \Omega \rightarrow \Omega$ that is a diffeomorphism of the ambient space. We used the computationally more efficient stationary velocity field (SVF) parametrisation of diffeomorphisms instead of the variable velocity fields or shooting formulations seen in LDDMM. The SVF preserves the same robustness guarantees as its alternatives \cite{ashburner2007fast} and is formally written as an initial value problem of the ordinary differential equation;
\begin{equation}
\frac{d\phi_t}{dt}=\textbf{v}(\phi_t), \text{with } \phi_0=Id.
\end{equation}
This is solved over unit time using a numerical integrator, resulting in the final deformation $\varphi = \phi_1^{v}$. Mathematically, the transformation $\varphi$ becomes the Lie group exponential map parametrised by $v$, $\varphi = Exp(v)$. In this work, we used an Euler integration with scaling and squaring similar to \cite{arsigny2006log}.
\begin{figure}[t]
  \centering
  \includegraphics[width=\linewidth]{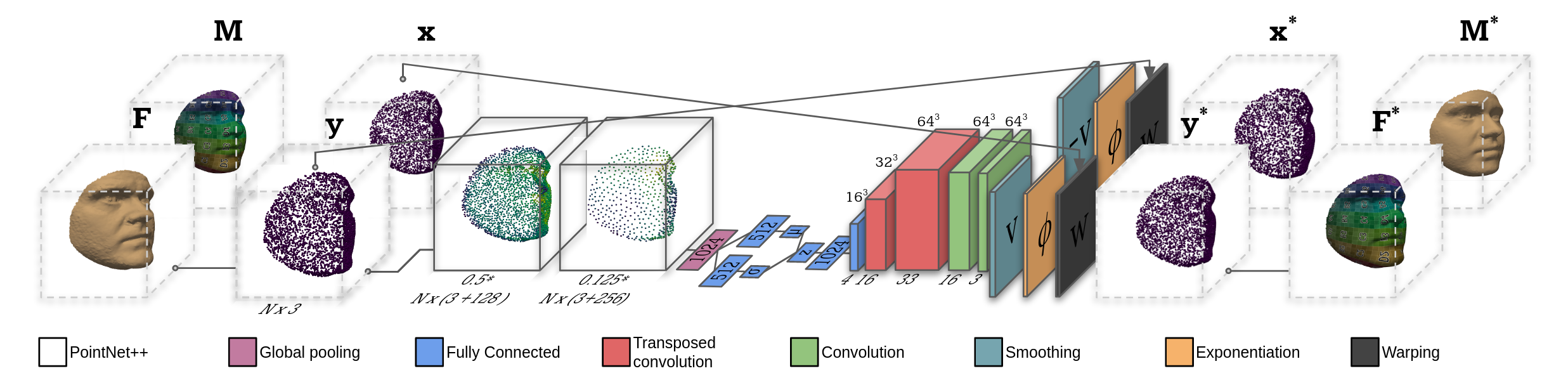}
  \caption{Network architecture}
  \label{fig:architecture}
\end{figure}
\subsection{CVAE Network}
We propose a CVAE \cite{kingma2014semi} that internalises a PDM, while being able to perform a one-step warp of the moving template surface to a fixed target surface. As input for the network, we used independently and randomly sampled point clouds $x$ and $y$ of the affine aligned moving ($M$) and fixed ($F$) mesh respectively, where $x, y \in \mathbb{R}^{N \times d}$. Point clouds for every surface were resampled at every epoch during training to improve robustness. A schematic representation of the model architecture is shown in figure \ref{fig:architecture}. The encoder predicts a probabilistic parametrisation for the deformation of $x$ given $y$. Given the fixed point cloud, the parametric encoder $g_\omega(\mu, \sigma;y)$ predicts the mean $\mu \in \mathbb{R}^Z$ and covariance $\sigma \in \mathbb{R}^Z$ of the posterior normal distribution $q_\omega(z|y) = \mathcal{N}(\mu, \sigma)$. Given the approximate posterior, an encoding $z\in\mathbb{R}^Z$ is sampled for the input. To structure the latent space, we made the approximate posterior distribution approach a prior distribution $p(z)=\mathcal{N}(0,I)$ using the Kullback-Leibler (KL) divergence. The decoder reconstructs point cloud $y$ by warping the conditional point cloud $x$ according to the predicted deformation parametrised by $z$, this results in the distribution $p_\gamma(y | x, z)$. The decoder is comprised of four blocks. First, the latent representation is decoded as a vector field. Second, the vector field is smoothed using a Gaussian smoothing layer, resulting in a velocity field $v\in \mathbb{R}^{V^d \times d}$. Third, the velocity field is integrated as explained in section \ref{subsec:reg_model} using an exponentiation layer \cite{dalca2018unsupervised, krebs2018unsupervised}, resulting in the final deformation function $\phi^v_1=Exp(v)$. Last, the moving and fixed point clouds are warped using the warp layer $y^* = \phi^v_1 \circ x$. Because we deform the ambient space instead of individual points, we can apply this same mapping to the vertices of the surface $M$ during inference and it allows us to choose the size of point clouds $x$ and $y$ independent of the mesh resolution.  

\subsection{Objective Function}
The network parameters $\theta = (\omega, \gamma)$ were trained end-to-end, minimising the variational lower bound; $ - \mathbb{E}_{z\sim q_{\omega}(z|y)} \log p_\gamma(y | x, z) +  KL(q_\omega(z|y) || p(z))$, where the first term is the reconstruction log-likelihood between the sampled points and the second term a divergence between the approximate posterior and prior. KL divergence can be computed in closed form \cite{kingma2014semi} and minimising the negative log-likelihood is equivalent to minimising the mean squared error (MSE):
\begin{equation}
    \mathcal{L}_{MSE}(x, y) = \frac{1}{N} \sum_{i=1}^N \lVert x_i - y_i \rVert ^2_2.
    \label{eq:mse}
\end{equation}
However, the MSE can only be computed if correspondence between shapes is known, which is rarely the case in real-world scenarios. Therefore, we experimented with two alternative loss functions. In a first instance, the Chamfer distance (CD) takes the closest point instead of the corresponding point:
\begin{equation}
\mathcal{L}_{CD}(x,y)=\frac{1}{N} \sum_{i=1}^N \underset{c \in y}{\min} \lVert x_i - c \rVert_2^2 + \frac{1}{N} \sum_{j=i}^N  \underset{c \in x}{\min} \lVert c - y_i \rVert_2^2.
    \label{eq:cd}
\end{equation}
As explained in \cite{feydy2020geometric}, the nearest neighbour projections of a CD tend to result in low quality gradients. Therefore, in a second and alternative instance, we also looked at optimal transport theory. Intuitively, a Wasserstein distance attempts to find the transport plan $\pi$ that orders the points. We approximated the Wasserstein distance using the debiased Sinkhorn divergence (SD) \cite{cuturi2013sinkhorn}
\begin{equation}
    \mathcal{L}_{SD}(\alpha, \beta, \epsilon)=OT_\epsilon(\alpha,\beta)-\frac{1}{2}OT_\epsilon(\alpha, \alpha)-\frac{1}{2}OT_\epsilon(\beta, \beta),
\label{eq:sd}
\end{equation}
where $OT_{\epsilon}$ the optimal transport distance with an entropic regularisation term. 
\begin{equation}
OT_{\epsilon}(\alpha, \beta)=\underset{\pi \in \Pi}{\min} \sum_{i=1}^N\sum_{j=1}^N \pi_{i,j} \frac{1}{p}\lVert  x_i - y_j \rVert_p + \epsilon KL(\pi \lVert \alpha \otimes \beta),
\end{equation}
subject to $\pi > 0, \pi\textbf{1}=\alpha, \pi^T\textbf{1}=\beta$, given $\pi$ the regularised transport plan, \textbf{1} vector of ones, and, $\alpha=\sum_{i=1}^N\alpha_i\delta_{x_i}$, $\beta=\sum_{j=1}^M\beta_j\delta_{y_j}$ probability measures. 
The SD approaches the Wasserstein distance as $\epsilon \rightarrow 0$. In this work, we used $\epsilon=10^{-4}$, $p=1$ and the geometric loss python package \cite{feydy2019interpolating}. The SD was first applied to diffeomorphic registration in \cite{feydy2017optimal}. 

Experimentally we have found that these constraints were not enough, resulting in unnecessarily complicated deformation fields and loss of correspondence. To overcome these problems, we introduced two additional regularisation terms. The first term encourages minimal deformation, the second term encourages points of the deformed shape and the original shape to stay close together, preventing drift of the landmarks:
\begin{align}
\mathcal{R}_{smooth}(v) = \frac{1}{3V^3}\sum_{i=1}^{3V^3}(-\alpha \nabla^2 v+ \gamma v)^2, && \mathcal{R}_{vert}(x, x^*) = \frac{1}{N} \sum_{i=1}^{N} \lVert x_i - x^*_i \Vert _{1}.
\end{align}
with $\nabla^2$ the Laplacian operator approximated using central finite differences, $\alpha=10^{-6}$ and $\gamma=1$. The final objective function is defined as: 
\begin{equation}
    \begin{split}
\mathcal{O}(x,y,z,v) = 
&\frac{\lambda_1}{2} (\mathcal{L}(y, \phi^{v}_1 \circ x) + \mathcal{L}(x, \phi^{-v}_0 \circ y)) + \lambda_2 \mathcal{R}_{smooth}(v) \\
&+ \frac{\lambda_3}{2} (\mathcal{R}_{vert}(x, \phi^{v}_1 \circ x) + \mathcal{R}_{vert}(y, \phi^{-v}_0 \circ y)) + \frac{1}{Z}KL(q_\omega(z|F) \lVert p(z)). 
\end{split}
\end{equation}
We found $\lambda_2=10$ and $\lambda_3=100$ to work best. $\lambda_1$ varies per loss function; $4\times10^4$, $8\times10^4$ and $5\times10^3$ for MSE, CD and SD respectively. Once the order of magnitude was found for these parameters we did not experience major differences in performance when adjusting the factor.

\section{Experiments and Results}
As dataset we used ``3D Facial Norms'' \cite{weinberg20163d}, which contains 3D facial surface scans of 2454 individuals (952M/1502F), the download is available through FaceBase \cite{FACEBASE:VWP}, a controlled-access data respository managed by the U.S. National Institutes of Health. We excluded 47 unsuitable scans due to imaging artifacts. The other images were cleaned by manually removing the clothing and hair, pose normalised with a 3D facial template based on 5 manually indicated landmarks followed by a rigid ICP procedure as implemented in the open-source toolbox MeshMonk \cite{white2019meshmonk}, and isotropically scaled to fit within a centered cube with side=2. The dataset was split in a training, validation and test set of sizes 2007, 200 and 200 respectively. 

We designed two experiments to examine different properties of the network in figure \ref{fig:architecture}, the same split was used in every experiment. In the first experiment (subsection \ref{subsec:reg}), the model was evaluated as a registration algorithm on the dataset without correspondence. As an additional preprocessing, we algoritmically cropped the meshes (removing e.g. extended neck or forehead) by finding the nearest vertex on $M$ (template) for every vertex on $F$ (facial surface) and removing the vertices that mapped to the border of $M$. We experimented with two loss functions (CD and SD) to cope with the lack of correspondence. In the second experiment (subsection \ref{subsec:pdm}), the model was evaluated as a PDM with the data brought into correspondence using MeshMonk \cite{white2019meshmonk}. This allowed us to exclude uncertainty of the loss function.

\subsection{Implementation Details}
We worked on $d=3$ dimensional space, with point clouds of $N = 5000$ samples. When spatial correspondence was known, the point cloud was obtained by uniformly sampling the same vertices from moving and fixed shape; when it was not known, point cloud sampling was performed uniformly taking mesh triangle area into account. The encoder $g_{\omega}$ consisted of two PointNet++ \cite{qi2017pointnet++} blocks as implemented in \cite{Fey_Lenssen_2019}. PointNet++ layers consist of a sampling grouping and PointNet layer and allow for hierarchical learning on point clouds \cite{qi2017pointnet++}. During sampling, we iteratively sampled a ratio of 0.5 and 0.25 of the furthest points for the first and second layer respectively. Grouping was performed in a ball query with a radius of 0.2 and 0.4 and a maximum of 64 neighbours. For the PointNet layer, we used 3 linear blocks, consisting of a linear, ReLU and batch normalisation layer of $3 \rightarrow 64 \rightarrow 64 \rightarrow 128$ channels in the first layer and $(3+128) \rightarrow 128 \rightarrow 128 \rightarrow 256$ channels in the second layer. The learned hierarchical representations are shown together with the model architecture in figure \ref{fig:architecture}. The features and positions are concatenated and fed in 3 similar linear blocks of $(256+3) \rightarrow 256 \rightarrow 512 \rightarrow 1024$ channels. This was followed by a global pooling layer, taking the maximum across the node dimensions, resulting in a 1024 dimensional feature vector. The bottleneck consisted of two pairs of two fully connected layers with an ELU and linear activation function resulting in $\sigma$, $\mu$. Due to memory constraints only a single latent encoding $z$ of predetermined size $Z$ was sampled per epoch from $q_\omega$. The latent encoding was decoded using two fully connected layers, two transposed convolution layers (4 kernel size, 1 padding and 2 stride), and two convolution layers (1 kernel size, 0 padding and 1 stride). Each layer was followed by an ELU activation function, the transposed convolutions were followed by a batch normalisation layer. The Gaussian smoothing layer uses a kernel of size 15 and standard deviation of $4$, the predicted vector field consisted of $V=64$ 3-dimensional vectors in every dimension. The exponentiation layer made use of $T = 7$ scaling and squaring steps. For the warp layer, we used trilinear interpolation to interpolate the deformation of every point from the deformation field. All CVAE models were trained for 300 epochs with batch size 7 using Adam optimizer with a learning rate of $10^{-5}$, the model using SD was trained with a batch size 3 due to memory constraints.
\begin{figure}[t]
\includegraphics[width=\textwidth]{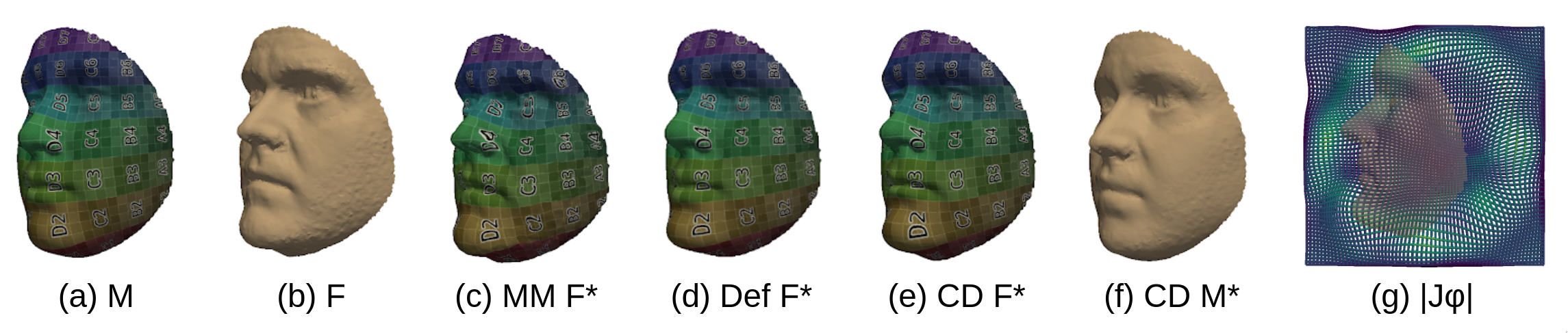}
\centering
\caption{Registration example on unseen data sample: a,b) moving shape, fixed shape; c) registration result of baseline MeshMonk (MM); d) registration result of Deformetrica (Def); e,f) registration result and inverse of CVAE using Chamfer distance (CD); g) slice 32 of deformation field color-coded by the Jacobian determinant (values range between 0.01 and 3.4).}
\label{fig:registration_result}
\end{figure}

\subsection{Validation of Registration} \label{subsec:reg}
\begin{figure}[t]
\centering
\begin{subfigure}{.25\textwidth}
  \centering
  \includegraphics[width=1\linewidth]{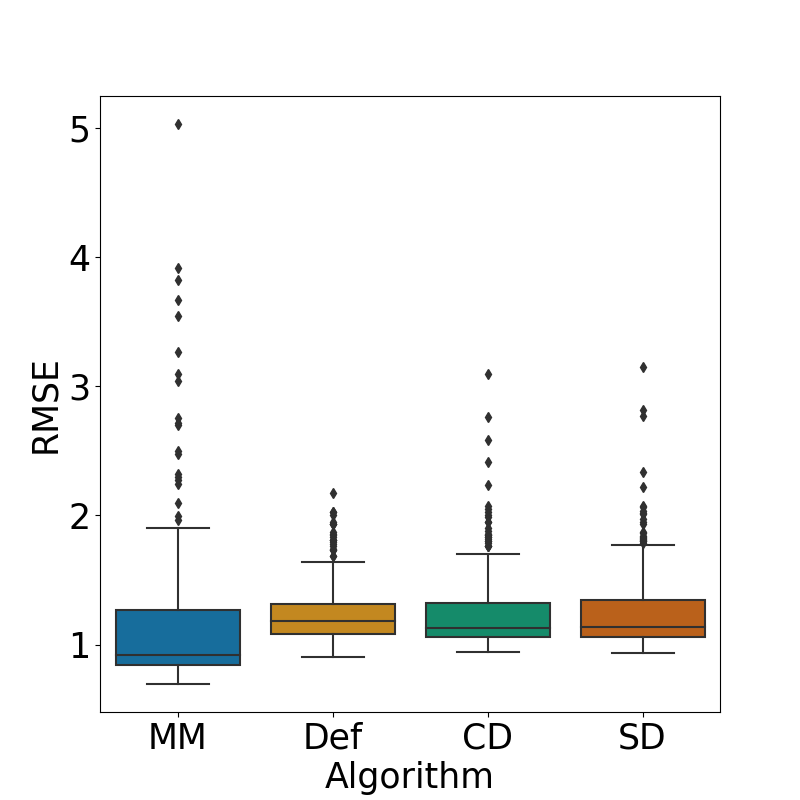}
  \caption{Fit}
  \label{subfig:fit}
\end{subfigure}%
\begin{subfigure}{.25\textwidth}
  \centering
  \includegraphics[width=1\linewidth]{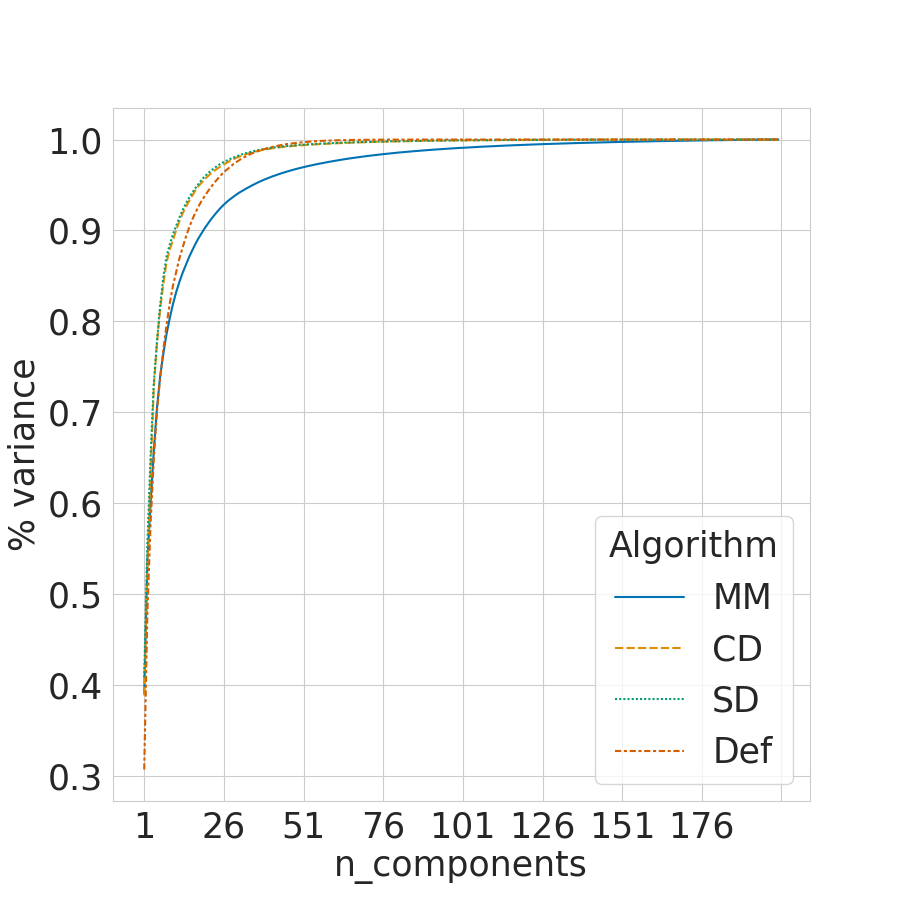}
  \caption{Compactness}
  \label{subfig:compactness}
\end{subfigure}%
\begin{subfigure}{0.25\textwidth}
  \centering
  \includegraphics[width=1\linewidth]{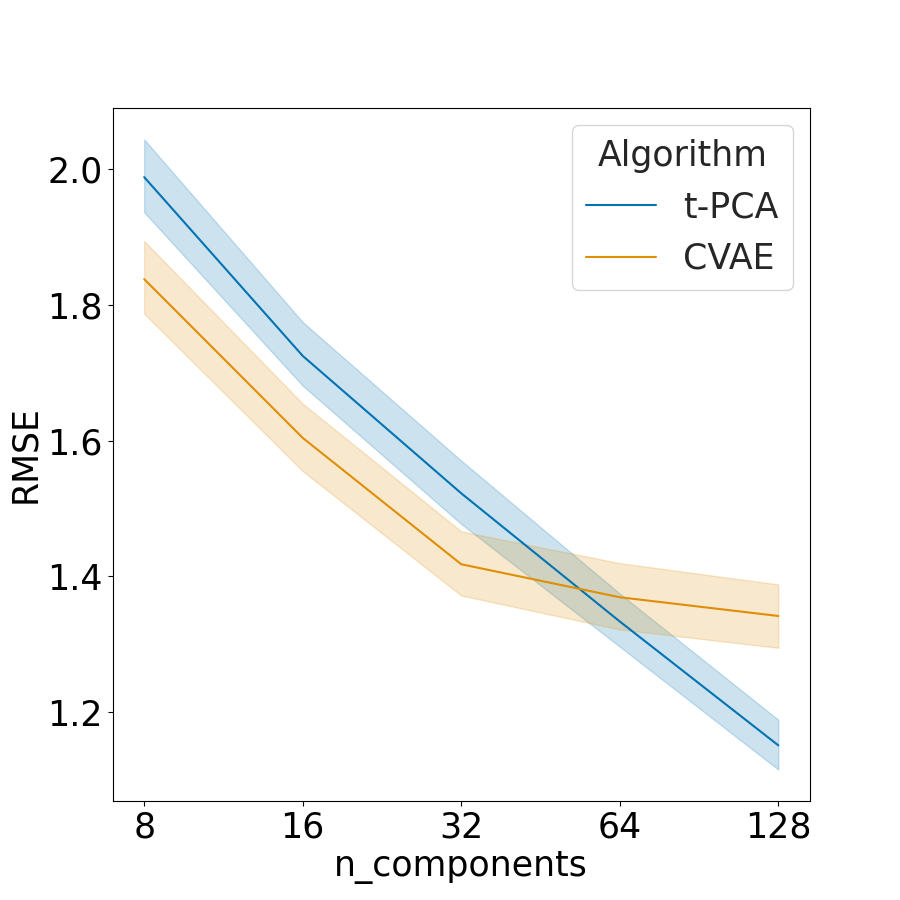}
  \caption{Generalisability}
  \label{subfig:generalisability}
\end{subfigure}%
\begin{subfigure}{.25\textwidth}
  \centering
  \includegraphics[width=1\linewidth]{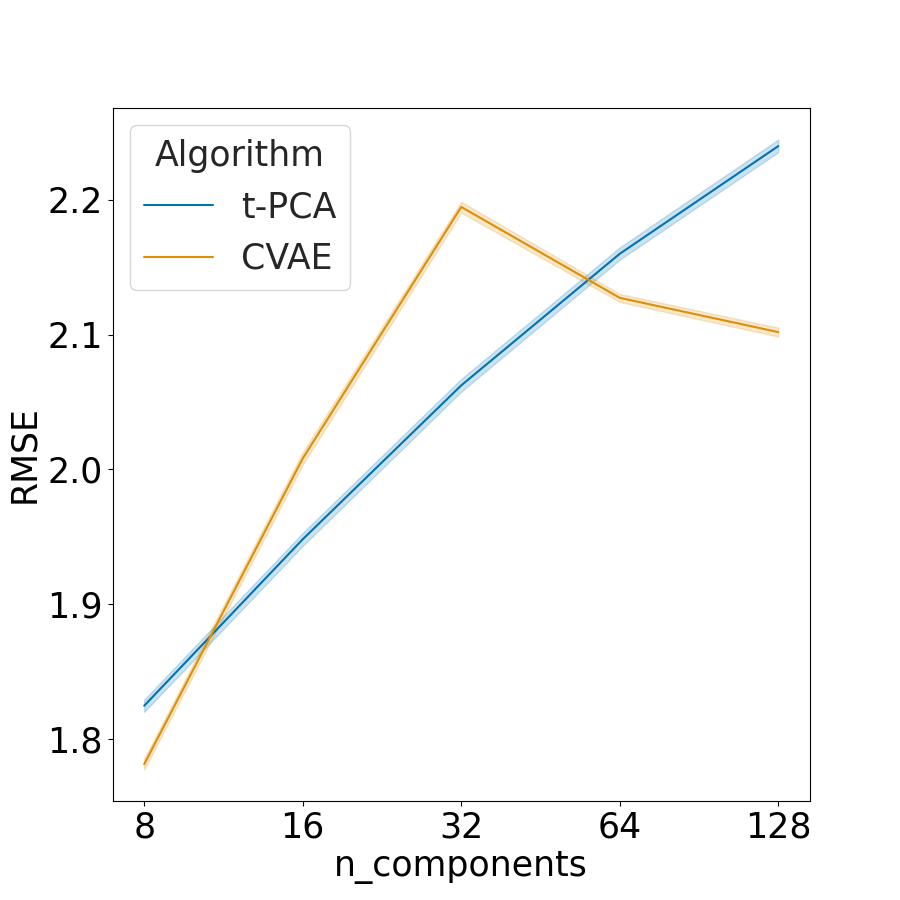}
  \caption{Specificity}
  \label{subfig:specificity}
\end{subfigure}
\caption{a,b) Registration performance: shape fit and compactness of baselines MeshMonk (MM) and Deformetrica (Def), and, CVAE trained with Chamfer distance (CD) and Sinkhorn divergence (SD); c,d) PDM performance: generalisability and specificity for CVAE and t-PCA on data in correspondence. }
\label{fig:spec_gen_n_components}
\end{figure}

In the first experiment, we evaluated the registration performance of the network using a latent size of $Z=32$. We compared the performances of the proposed model with CD (equation \ref{eq:cd}) and SD (equation \ref{eq:sd}) as distance function to MeshMonk (MM) \cite{white2019meshmonk} and Deformetrica (Def) \cite{bone2018deformetrica} as baseline. We used recommended parameters for both, in Def we used a varifold distance with $10^{-6}$ noise-std. A registration result of MM, Def and CD are displayed together in figure \ref{fig:registration_result}. The performance was quantified as shape fit and model compactness. The shape fit is the RMSE between $M^*$ and $F$. Because correspondence in this experiment was not known, we used the average squared error over the 3-nearest vertices as distance between vertices. Figure \ref{subfig:fit} shows a median RMSE by MeshMonk of $0.92$, better than $1.18$ for Def, $1.13$ for CD and $1.14$ for SD. For the Diffeomorphic techniques, the distributions of errors in shape fit tend to be closer to the median, this could be a direct consequence of the Diffeomorphic constraints. However, for the CVAE it could also be due to the internal PDM, which acts as regularisation. This same constraints could also explain the slightly lower accuracy in terms of shape fit. To validate the indication of corresponding points we looked at model compactness. In the context of surface registration, a more compact shape model typically originates from an improved (more consistent) indication of corresponding points \cite{davies2002minimum}. This was quantified as the cumulative percentage of explained variance of a PCA point-distribution model on the correspondences outputted by the registration pipeline. Figure \ref{subfig:compactness} shows a higher compactness with an AUC of $195.29$ and $195.48$ for CD and SD respectively, compared to $192.01$ for MM and $194.42$ for Def. This potentially indicates an improved quality of correspondences as obtained by the proposed models. However, one should always take into account the trade-off between model compactness and goodness-of-fit (and its dependence on hyperparameter choices). If the CVAE results in shapes that are closer to the template we would indeed expect a lower model complexity. Finally, we did not observe significant differences between SD and CD; this is probably due to faces already being very similar in shape. Inference took on average 0.1s on NVIDIA GeForce RTX 2080 Ti. 

\subsection{Validation of Internal Probabilistic Deformation Model} \label{subsec:pdm}
In the second experiment, we additionally evaluated the performance of the proposed model as a non-linear PDM given spatially dense correspondences. Having correspondence allowed us to use exact point-to-point distances, excluding the uncertainty of shape fit, focusing solely on the degrees of freedom offered by the PDM and regularisation terms. We used the MSE (equation \ref{eq:mse}) as distance metric. As baseline we used iterative optimisation to find the training set SVFs
\begin{equation}
v = \underset{v}{\arg\max} \frac{\lambda_1}{2} (\mathcal{L}_{MSE}(y, \phi^{v}_1 \circ x) + \mathcal{L}_{MSE}(x, \phi^{-v}_0 \circ y)) + \lambda_2 \mathcal{R}_{smooth}(v),
\end{equation}
followed by incremental PCA \cite{ross2008incremental} on the velocity fields (incremental PCA has better memory scaling than traditional PCA), this is referred to as tangent-PCA (t-PCA). Similar to the CVAE, the velocity fields were also smoothed using the same Gaussian kernel. The first 5 components for both t-PCA and the CVAE are visualised in supplementary material A, figure 1. The performance of the deformation model was quantified as a trade-off between generalisability and specificity in terms of the shape it produced. In statistical shape analysis, generalisability is the ability of a model to fit to unseen data, measuring the RMSE between $M^*$ and $F$ on the test set. Figure \ref{subfig:generalisability} shows a better generalisability for the CVAE using a lower number of components $\leq32$ and an earlier plateauing compared to the t-PCA baseline. Specificity is the ability to generate shapes that are similar to the training set, measured by sampling the latent space according to its distribution and averaging the RMSE of the generated example to the 3-nearest neighbour shapes in the training set. Here we used 10,000 generated shapes. Figure \ref{subfig:specificity} shows an upward trend as the model gets more freedom to express the entire deformation space. A drop in specificity for the proposed model is observed where the number of components $>32$. This could be due to $\mathcal{R}_{vert}$ over-constraining the model, however, more experimentation is required.

\section{Discussion and Conclusion}
In this work, we proposed an unsupervised end-to-end one-step surface registration technique that jointly learns a probabilistic deformation model. The internalised PDM is competitive to linear PCA and registration results are competitive to iterative non-rigid registration algorithms that are widely used in the field. Furthermore because this technique deforms the ambient space and we work with a latent encoding the technique can easily be generalised to other image representations, showing potential for real time hybrid shape and deformation analysis software. A limitation of this work is that the parameters have been specifically fine tuned for face registration, applying it to a different anatomical structure would require retraining. Another limitation is the lack of ground truth data, which makes evaluation challenging. A last limitation is that this network is restricted to a single moving template shape. Future work involves replacing the vector field by control points, incorporating surface to volume registration and performing statistics on the predicted deformations. In conclusion, we propose a registration algorithm that, in contrast to traditional techniques, does not require iterative optimisation; is guaranteed to be diffeomorphic; works on any mesh resolution; and internalises deformation constraints. The performance of the proposed model is competitive to traditional techniques and can easily be integrated in other pipelines.\\

\noindent\textbf{Acknowledgements.} This project was funded by the Research Foundation in Flanders (FWO, Fonds Wetenschappelijk Onderzoek) Phd Fellowship (1SB0121N), post-doctoral fellowship (12ZV420N) and research project (G078518N), by the Research Fund KU Leuven (BOF-C1, C14/15/081 and C14/20/081) and by ERC Consolidator grant No. 724228 (LEMAN).

%
%
%
\bibliographystyle{splncs04}
\bibliography{references}

\end{document}